\documentstyle[amssymb,prd,aps,multicols,epsf]{revtex}

\begin{document}
\title{Oscillatons revisited}
\author{L. Arturo Ure\~{n}a-L\'{o}pez\thanks{E-mail address: lurena@fis.cinvestav.mx}}
\address{Departamento de F\'{\i}sica, \\
Centro de Investigaci\'on y de Estudios Avanzados del IPN,\\
AP 14-740, 07000 M\'exico D.F., MEXICO.\\
}
\date{\today}
\maketitle

\begin{abstract}
In this paper, we study some interesting properties of a spherically symmetric oscillating soliton star made of a real time-dependent scalar field which is called an {\it oscillaton}. The known final configuration of an oscillaton consists of a stationary stage in which the scalar field and the metric coefficients oscillate in time if the scalar potential is quadratic. The differential equations that arise in the simplest approximation, that of coherent scalar oscillations, are presented for a quadratic scalar potential. This allows us to take a closer look at the interesting properties of these oscillating objects. The leading terms of the solutions considering a quartic and a cosh scalar potentials are worked in the so called stationary limit procedure. This procedure reveals the form in which oscillatons and boson stars may be related and useful information about oscillatons is obtained from the known results of boson stars. Oscillatons could compete with boson stars as interesting astrophysical objects, since they would be predicted by scalar field dark matter models.
\end{abstract}

\draft

\begin{multicols}{2}  
\narrowtext

\section{Introduction}

Being the nature of dark matter one of the most intriguing open questions in physics nowadays\cite{fur,stein,salucci,cabral}, it is not strange that the number of proposals trying to solve it by considering exotic matter has increased in recent years. Among them, models of bosonic dark matter seem promising and have been widely studied, both at cosmological and galactic scales\cite{seidel91,tkachev,peebles,hu,bento,sahni,matos,matos2,luis1,luis2,luis3,matos3,matos4,matos5,lidsey,colpi,pang,seidel90,sin,seidel94,seidel98,diego2,diego,arbey}. On one hand, complex scalar field boson stars are the most studied as objects of astrophysical interest, see for instance\cite{colpi,pang,seidel90,sin,seidel94,seidel98,diego2,diego,arbey} and many references there in. On the other hand, bounded objects formed by real scalar fields has not been studied that deeply yet\cite{seidel91,tkachev,peebles,hu,matos,matos2,matos3,matos5,seidel94,becerril}.

The study of gravitationally bound objects with real scalar fields is a very important issue for the scalar field dark matter hypothesis at galactic scales\cite{matos3,matos4,matos5}, but most of the recent treatments have been static, both Newtonian and relativistic\cite{tkachev,peebles,hu,matos,matos2}. It is known that the only fully relativistic bound object made of real scalar field, the oscillaton, is time-dependent in nature\cite{seidel91,matos3,matos5,seidel94}, in opposition to the static treatments that can be made for boson stars, i.e., for complex scalar fields. Besides the possible interest for cosmologists, from the point of view of general relativity, oscillatons are also important since they are time dependent, non-topological, non singular and asymptotically flat solutions to the coupled Einstein-Klein-Gordon equations. Their intrinsic properties are quite different than those of the known static, or even stationary, solutions to the Einstein equations. Moreover, their stability and possible formation by means of a Jeans mechanism has been studied partially before\cite{seidel91,matos5,seidel94}, but more studies are needed in order to establish their possible astrophysical role.

In this paper, our main aim is to make a generic study of the solutions to the spherically symmetric coupled Einstein-Klein-Gordon equations for a time-dependent real scalar field $\Phi$ endowed with a quadratic potential, using a simple approximation. Within this approximation, we can manipulate the differential equations, solve them numerically and obtain, easily and clearly, some of the interesting properties of oscillatons. This elucidates the results presented in previous papers about these objects. Also, we show how we can extend this study to other scalar potentials, which have not been treated before but that have been taken into account in scalar field dark matter models.

A brief summary of the paper is as follows. In section II, we show how to obtain the equations for an oscillaton with a quadratic scalar potential, under the assumption of coherent scalar field oscillations with fundamental frequency $\omega$, that is, the scalar field is of the form $\Phi(t,r)=\sigma(r) \cos (\omega t)$. While the general form of the metric coefficients can be known within this assumption, an approximation in Fourier series is done in order to deal with the full differential equations. It is discussed in turn whether this approximation is enough, even though the Klein-Gordon equation for the scalar field has a time-dependent damping term. Taking appropriate dimensionless quantities and boundary conditions, we find some non singular and asymptotically flat 0-node solutions. Since the solutions are also asymptotically static, in the sense that the oscillations of the oscillaton are confined to a finite spatial region, the mass of the configurations is the Schwarzschild mass that an observer measures at infinity. This allows us to calculate the critical mass of oscillatons.

In section III, the similarity between a boson star and an oscillaton is discussed under the so called stationary limit procedure. This limit can be seen as the lowest order of approximation to an oscillaton in which the differential equations become boson star-like. Some numerical results are presented for comparison between an oscillaton and a boson star. The similarity is better for weaker fields and shows that oscillatons can become interesting astrophysical objects at the same level of boson stars.

The case of a scalar field with a quartic self-interaction is worked within the stationary limit procedure in section IV (A). The resulting equations can also be written in a boson star-like form, which permits us to easily obtain information for oscillatons using the known results of boson stars. This limit is also applied to the case of a (non-polynomial) cosh-potential in section IV (B), but the resulting equations are not boson star-like. However, we can use the similarity between cosh and quartic oscillatons to obtain useful information. In section V, we summarize the results and give points to be investigated in future research. The numerical method used in this paper is the fourth-order Runge-Kutta method of {\small MAPLE}.

\section{Oscillatons} 

We start by taking the case of a quadratic scalar interaction with spherical symmetry, that was analyzed in\cite{seidel91}. The most general spherically-symmetric metric is written as

\begin{equation}
ds^2=g_{\alpha \beta} dx^\alpha dx^\beta = -e^{\nu-\mu} dt^2 + e^{\nu+\mu} dr^2 + r^2 d\Omega^2
\end{equation}

\noindent where $\nu=\nu(t,r)$, $\mu=\mu(t,r)$ are yet arbitrary functions (the units are such $c=1$). The energy-momentum tensor of a real scalar field $\Phi(t,r)$ endowed with a scalar field potential $V(\Phi)$ is defined as\cite{wald}

\begin{equation}
T_{\alpha \beta} = \Phi_{,\alpha} \Phi_{,\beta} - \frac{1}{2} g_{\alpha \beta} \left[ \Phi^{,\lambda} \Phi_{,\lambda} + 2 V(\Phi) \right], \label{tensor}
\end{equation}

\noindent whose non-vanishing components are

\begin{eqnarray}
-{T^0}_0 &=& \rho_\Phi = \frac{1}{2} \left[ e^{-(\nu-\mu)} {\dot{\Phi}}^2 + e^{-(\nu+\mu)} \Phi^{\prime 2} + 2 V(\Phi) \right] \label{tensor1} \\
T_{01} &=& {\mathcal P}_{\Phi} = \dot{\Phi} \Phi^\prime \label{tensor2} \\
{T^1}_1 &=& p_r = \frac{1}{2} \left[ e^{-(\nu-\mu)} {\dot{\Phi}}^2 + e^{-(\nu+\mu)} \Phi^{\prime 2} - 2 V(\Phi) \right] \label{tensor3} \\
{T^2}_2 &=& p_\bot = \frac{1}{2} \left[ e^{-(\nu-\mu)} {\dot{\Phi}}^2 - e^{-(\nu+\mu)} \Phi^{\prime 2} - 2 V(\Phi) \right] \label{tensor4}
\end{eqnarray}

\noindent and also ${T^3}_3 = {T^2}_2$. Overdots denote $\partial/\partial t$ and primes denote $\partial/\partial r$. These different components are identified as the energy density $\rho_\Phi$, the momentum density ${\mathcal P}_\Phi$, the radial pressure $p_r$ and the angular pressure $p_\bot$.

The Einstein equations, $G_{\alpha \beta}= \kappa_0 T_{\alpha \beta}$, can be written as differential equations for the functions $\nu,\mu$, then

\begin{eqnarray}
(\nu+\mu)^{\cdot} &=& \kappa_0 r \dot{\Phi} \Phi^\prime, \label{e1} \\
\nu^\prime &=& \frac{\kappa_0 r}{2} \left( e^{2\mu} \dot{\Phi}^2 + {\Phi^\prime}^2 \right), \label{e2} \\
\mu^\prime &=& \frac{1}{r} \left[1+e^{\nu+\mu}\left(\kappa_0 r^2 V-1 \right) \right], \label{e3}
\end{eqnarray}

\noindent where $\kappa_0 =8\pi G= 8\pi/m^2_{Pl}$ is the inverse of the reduced Planck mass squared. The conservation equations for the scalar field energy-momentum tensor (\ref{tensor}) are written as

\begin{equation}
{T^{\alpha \beta}}_{;\beta} = \Phi^{,\alpha} \left( \Box \Phi - \frac{dV}{d\Phi} \right)= 0, \label{kgc}
\end{equation}

\noindent where $\Box=g^{\alpha \beta} \nabla_\alpha \nabla_\beta$ is the d'Alambertian operator. Therefore, we obtain the Klein-Gordon (KG) equation for the scalar field $\Phi$,

\begin{equation}
\Phi^{\prime \prime} +\Phi^\prime \left(\frac{2}{r}-\mu^\prime \right) - e^{\nu+\mu} \frac{dV}{d\Phi} = e^{2\mu} \left( \ddot{\Phi} + \dot{\mu} \dot{\Phi} \right). \label{kg}
\end{equation}

Observe that the metric function $\nu$ is related to the scalar kinetic energy while $\mu$ is related to the scalar potential energy $V(\Phi)$. Actually, eqs. (\ref{e1}-\ref{kg}) are easier than the usual ones\cite{seidel91}. Notice that the left-hand side of the KG equation has the usual form of the static case and then the right-hand side looks like a source term. From this, it is not surprising that one can get non-singular solutions.

The quadratic scalar potential is written as $V(\Phi)=m^2 \Phi^2/2$, where $m$ is the boson mass. If we choose $\Phi(t,r)=\sigma(r) \phi(t)$, eq. (\ref{kg}) can be rewritten as

\begin{equation}
\frac{e^{-2\mu} }{\sigma} \left[ \sigma^{\prime \prime} +\sigma^\prime \left(\frac{2}{r}-\mu^\prime \right) - e^{\nu+\mu} m^2 \sigma \right] = \frac{1}{\phi} \left( \ddot{\phi} + \dot{\mu} \dot{\phi} \right). \nonumber
\end{equation}

\noindent This equation is almost separable. The terms on the right-hand side suggest that the scalar field oscillates harmonically in time with a damping term related to $\dot{\mu}$. Following the work\cite{seidel91}, we just consider that 

\begin{equation}
\sqrt{\kappa_0} \Phi(t,r)= 2 \sigma(r) \cos{(\omega t)}, \label{sphi}
\end{equation}

\noindent where $\omega$ is the frequency of the scalar oscillations. 

For simplicity, only this first approximation will be used along this paper, in order to analyze and reveal the main properties of oscillatons, as the procedure used in\cite{seidel91} was somewhat obscure. Besides, this is the most simple approximation since we are not including a damping term. I will discuss this at the end of this section. 

Eq. (\ref{e1}) can be formally integrated up to 

\begin{equation}
\nu + \mu = (\nu + \mu)_0 + r \sigma \sigma^\prime \cos{(2\omega t)}, \label{e1s}
\end{equation}

\noindent with $(\nu + \mu)_0$ an arbitrary function of the $r-$coordinate only. The metric functions $\nu$, $\mu$ should then be expanded as

\begin{eqnarray}
\nu (t,r) &=& \nu_0 (r) + \nu_1 (r) \cos{(2 \omega t)}, \nonumber \\
\mu (t,r) &=& \mu_0 (r) + \mu_1 (r) \cos{(2 \omega t)}. \label{numu}
\end{eqnarray}

\noindent Taking into account the Fourier expansion

\begin{equation}
e^{f(x) \cos{(2\theta)}} = I_0 (f(x)) + 2 \sum^\infty_{n=1} I_n (f(x)) \cos{(2n\theta)}, \label{fourier}
\end{equation}

\noindent where $I_n(z)$ are the modified Bessel functions of the first kind, we can expand the metric coefficients as 

\begin{eqnarray}
e^{\nu+\mu} &=& e^{\nu_0+\mu_0} \left[ I_0 \left( \nu_1+\mu_1 \right) + 2 \sum^\infty_{n=1} I_n \left(  \nu_1+\mu_1 \right) \cos{(2n \omega t)} \right], \nonumber \\
e^{\nu-\mu} &=& e^{\nu_0-\mu_0} \left[ I_0 \left( \nu_1-\mu_1 \right) + 2 \sum^\infty_{n=1} I_n \left( \nu_1-\mu_1 \right) \cos{(2n \omega t)} \right]. \label{metric}
\end{eqnarray}

\noindent The metric coefficients also oscillate in time but with even-multiples of the fundamental frequency $\omega$.

\subsection{Differential equations}

For numerical purposes, we perform a change of variables as in the boson star case\cite{pang,seidel90} 

\begin{eqnarray}
x=mr, &\,& \Omega = \frac{\omega}{m}, \nonumber \\
e^{\nu_0} \rightarrow e^{\nu_0} \Omega, &\,& e^{\mu_0} \rightarrow e^{\mu_0} \Omega^{-1}, \label{dimenless}
\end{eqnarray}

\noindent from which the metric coefficients are given now by $g_{rr}=e^{\nu+\mu}$ and $g_{tt}=-\Omega^2 e^{\nu-\mu}$. Observe that the boson mass sets both the scale of time and distance. 

The differential equations for the metric functions (\ref{numu}) and the scalar field (\ref{sphi}) appear from equations (\ref{e2}-\ref{kg}) by setting each Fourier component to zero. In contrast to the truncated expansion in\cite{seidel91}, we can use as much terms as necessary in expansions (\ref{metric}), i.e., there is no a priori truncation. The equations to be solved are

\begin{eqnarray}
\nu^\prime_1 &=& x \left[ e^{2 \mu_0} \sigma^2 \left( 2 I_1(2 \mu_1) - I_0(2 \mu_1) - I_2(2 \mu_1) \right) + {\sigma^\prime}^2 \right], \label{e2f2} \\
\nu^\prime_0 &=& x \left[ e^{2 \mu_0} \sigma^2 \left( I_0(2 \mu_1) - I_1(2 \mu_1)\right) + {\sigma^\prime}^2 \right], \label{e2f1} \\
\mu^\prime_0 &=& \frac{1}{x} \left\{ 1 + e^{\nu_0+\mu_0} \left[ x^2 \sigma^2 \left( I_0(x\sigma \sigma^\prime) + I_1(x\sigma \sigma^\prime) \right) \right. \right. \nonumber \\
&& \left. \left. - I_0(x\sigma \sigma^\prime) \right]\right\}, \label{e3f1} \\
\mu^\prime_1 &=& \frac{1}{x} e^{\nu_0+\mu_0} \left[ x^2 \sigma^2 \left( I_0(x\sigma \sigma^\prime) + 2 I_1(x\sigma \sigma^\prime) + I_2(x\sigma \sigma^\prime) \right) \right. \nonumber \\
&& \left. - \frac{}{} 2 I_1(x\sigma \sigma^\prime) \right], \label{e3f2} \\
\sigma^{\prime \prime} &=& - \sigma^\prime \left( \frac{2}{x} - \mu^\prime_0 -\frac{1}{2} \mu^\prime_1 \right) \nonumber \\
&&+ e^{\nu_0+\mu_0} \sigma \left( I_0(x\sigma \sigma^\prime) + I_1(x\sigma \sigma^\prime) \right)  \nonumber \\
&& - e^{2\mu_0} \sigma \left[ I_0(2\mu_1) \left( 1-\mu_1\right) + I_1(2\mu_1) + \mu_1 I_2(2\mu_1) \right], \label{kgf}
\end{eqnarray}

\noindent where now primes denote $d/d x$. 

In making the expansions (\ref{e2f2}-\ref{e3f2}), the neglected terms on the right hand side were those containing $\cos{(4\omega t)}$, while the neglected ones in the KG equation were those with $\cos (3\omega t),\, \cos{(5\omega t)}$. This suggests that the metric coefficients should be expanded with even Fourier terms, and that the scalar field expansion involves only odd Fourier terms. Then, the expansions used in\cite{seidel91} are well justified.

The solution is completely determined by solving eqs. (\ref{e2f2}-\ref{kgf}). But first, observe that eq. (\ref{e1}) gives the {\it exact algebraic} relation

\begin{equation}
\nu_1 + \mu_1 = x \sigma \sigma^\prime. \label{e1f}
\end{equation}

\noindent We will take this last equation to calculate function $\nu_1$ instead of integrating eq. (\ref{e2f2}), and then we are left with one algebraic and four ordinary differential equations.

\subsection{Boundary conditions}

In order to have nonsingular solutions, regularity at $x=0$ requires $\sigma^\prime (0)=0$ and $\nu (t,0)+\mu (t,0)=0$, so $\nu_0 (0) = - \mu_0 (0)$ and $\mu_1 (0)= - \nu_1 (0)$. But the latter condition is satisfied straightforwardly from eq. (\ref{e1f}). 

Asymptotic flatness require that $\sigma(\infty) = \mu_1 (\infty)=0$, which directly implies $\nu_1 (\infty)=0$. However, $\mu_0 (\infty)=-\nu_0 (\infty) \neq 0$ because of the change of variables in (\ref{dimenless}), and thus $\exp{(\nu-\mu)} (\infty) = \Omega^{-2}$, giving the value of the fundamental frequency $\omega$\cite{pang,seidel90}, but still $\exp{(\nu+\mu)} (\infty) = 1$. This also has the advantage that for each value of $\sigma (0)$, we only have two degrees of freedom and then we need only to adjust the central values $\mu_0 (0)$,$\mu_1(0)$ to obtain different n-node solutions. However, we will deal with 0-node solutions only.

The central value $\mu_1 (0)$ deserves special attention. From eq. (\ref{e3f2}), we see that its radial derivative is always positive. Thus, the asymptotically flat condition is only reached if one chooses $\mu_1 (0) < 0$. Also, the condition $|\mu_1| < 1$ is needed for the solutions of eqs. (\ref{e2f1}-\ref{kgf}) to converge, because we can neglect higher terms in expansion (\ref{metric}). As we will find later (see Fig. \ref{fig:metfs}), the central values $\sigma(0),\mu_0(0),\mu_1 (0)$ are of the same order of magnitude.

Needless to say, there is only one solution satisfying the above boundary conditions for each central value $\sigma(0)$. The procedure to solve numerically (\ref{e2f1}-\ref{e1f}) is similar as for boson stars.

\subsection{Numerical results}

\begin{figure}
\centerline{ \epsfysize=6cm \epsfbox{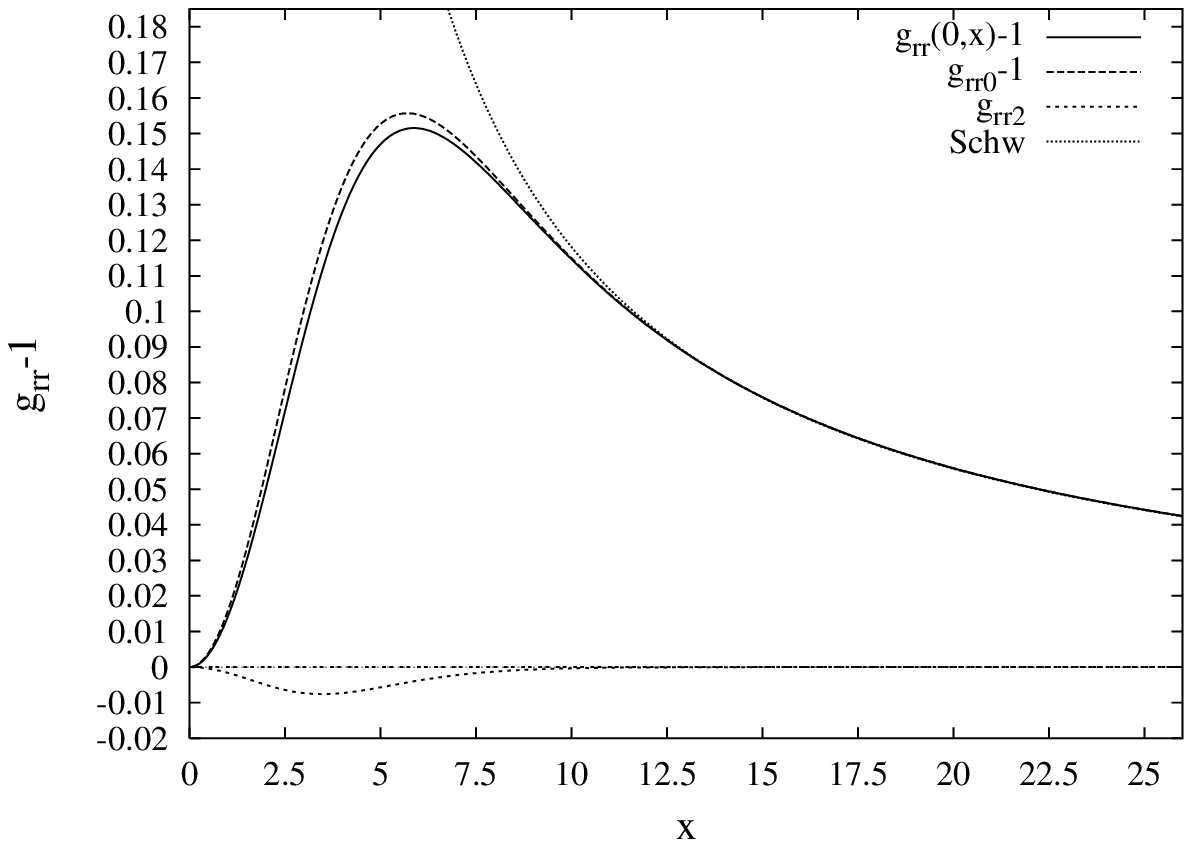}}
\centerline{ \epsfysize=6cm \epsfbox{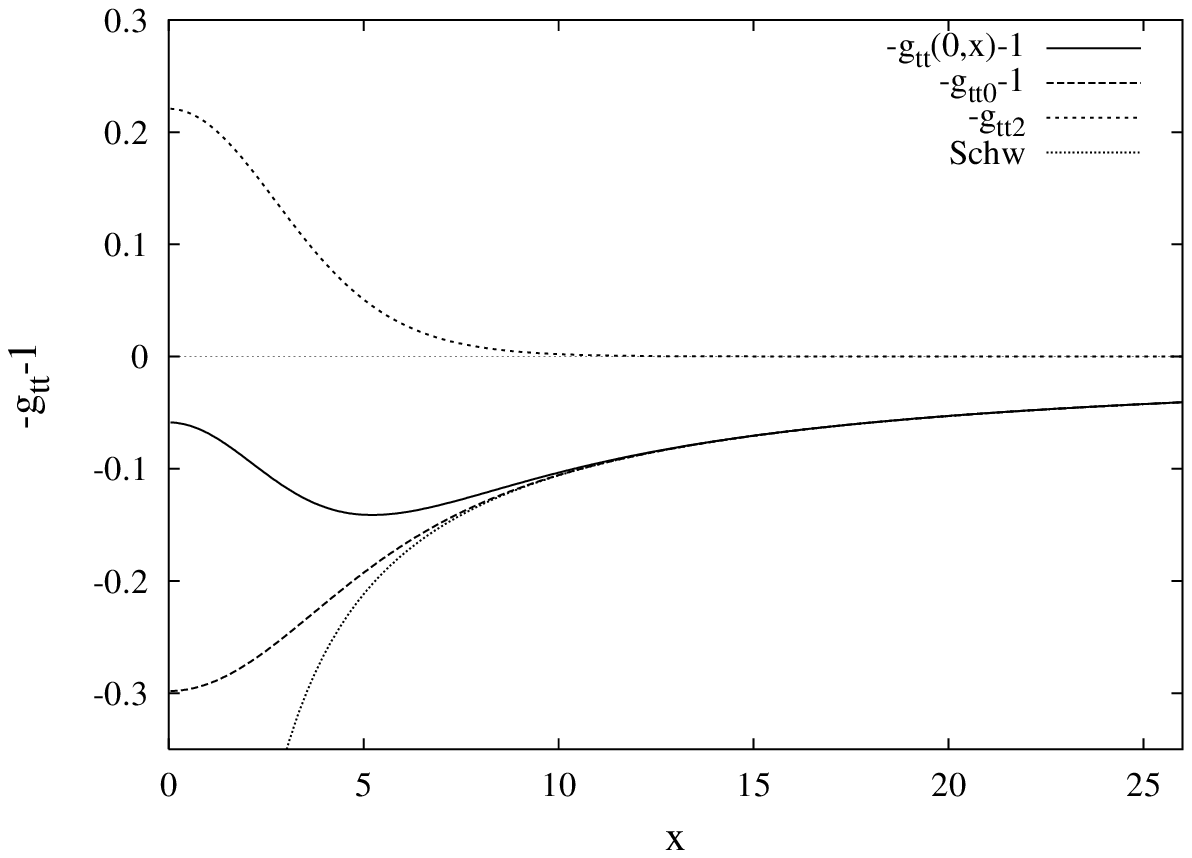}}
\caption{Metric coefficients $g_{rr}-1$ and $-g_{tt}-1$ and their respective two first terms in expansion, ${g_{rr}}_0-1$, ${g_{rr}}_2$ (top) and $-{g_{tt}}_0-1$, $-{g_{tt}}_2$ (bottom), as explained in the text for a central value $\sigma(0)=0.2/\sqrt{2}$. The mass of this 0-node oscillaton is $M=0.5295 \, m^2_{Pl}/m$ (see eq.~(\ref{mass}) and Fig.~\ref{fig:rhos}), and the Schwarzschild metric coefficients are also shown for a solution of the same mass. The calculated fundamental frequency is $\Omega=0.9183$ (see eq.~(\ref{freq}) and Fig.~\ref{fig:omegas}).}
\label{fig:metos}
\end{figure}

Typical metric coefficients for a 0-node solution with a central value $\sigma(0)=0.2/\sqrt{2}$ are shown in Fig. \ref{fig:metos}. We can identify the graphs in the figure by comparing the expansion

\begin{equation}
g=g_0(x) + g_2(x) \cos{(2\omega t)}+g_4(x) \cos{(4\omega t)} + ... \label{fexpg}
\end{equation}

\noindent with the corresponding expansion in (\ref{metric}) for each metric coefficient. The general behavior of the metric coefficients is similar to boson stars, see for example\cite{diego}.

\begin{figure}
\centerline{ \epsfysize=6cm \epsfbox{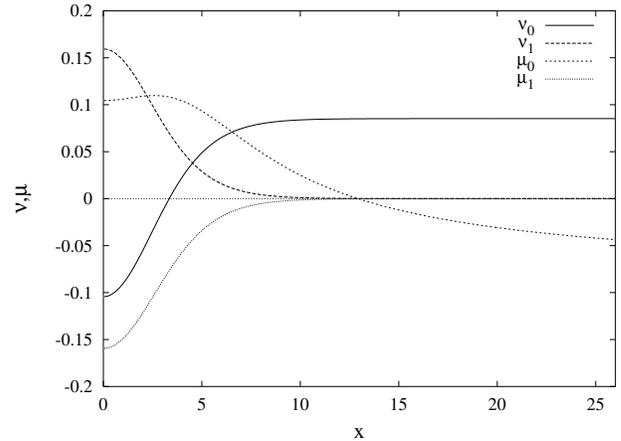}}
\caption{Metric functions $\nu_0$, $\nu_1$, $\mu_0$ and $\mu_1$ for the solution given in Fig. \ref{fig:metos}.}
\label{fig:metfs}
\end{figure}

The metric functions $\nu_0,\, \nu_1,\, \mu_0, \,\mu_1$ are shown in Fig. \ref{fig:metfs}. It can be verified that the required boundary conditions are satisfied. As we said before, the central values are of the same order than $\sigma(0)$ while the quantity $(\nu_1+\mu_1)$ is at least an order of magnitude smaller. These are typical behaviors.

\begin{figure}
\centerline{ \epsfysize=6cm \epsfbox{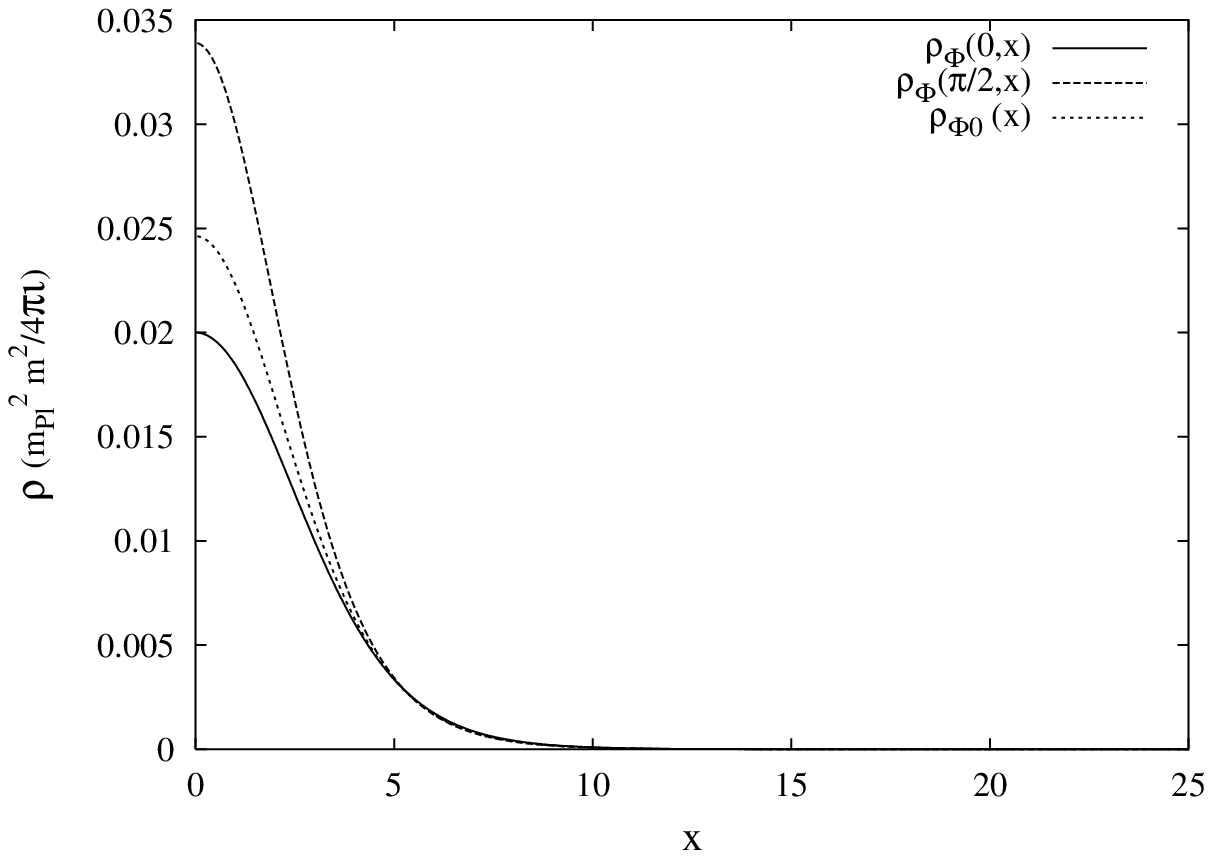}}
\centerline{ \epsfysize=6cm \epsfbox{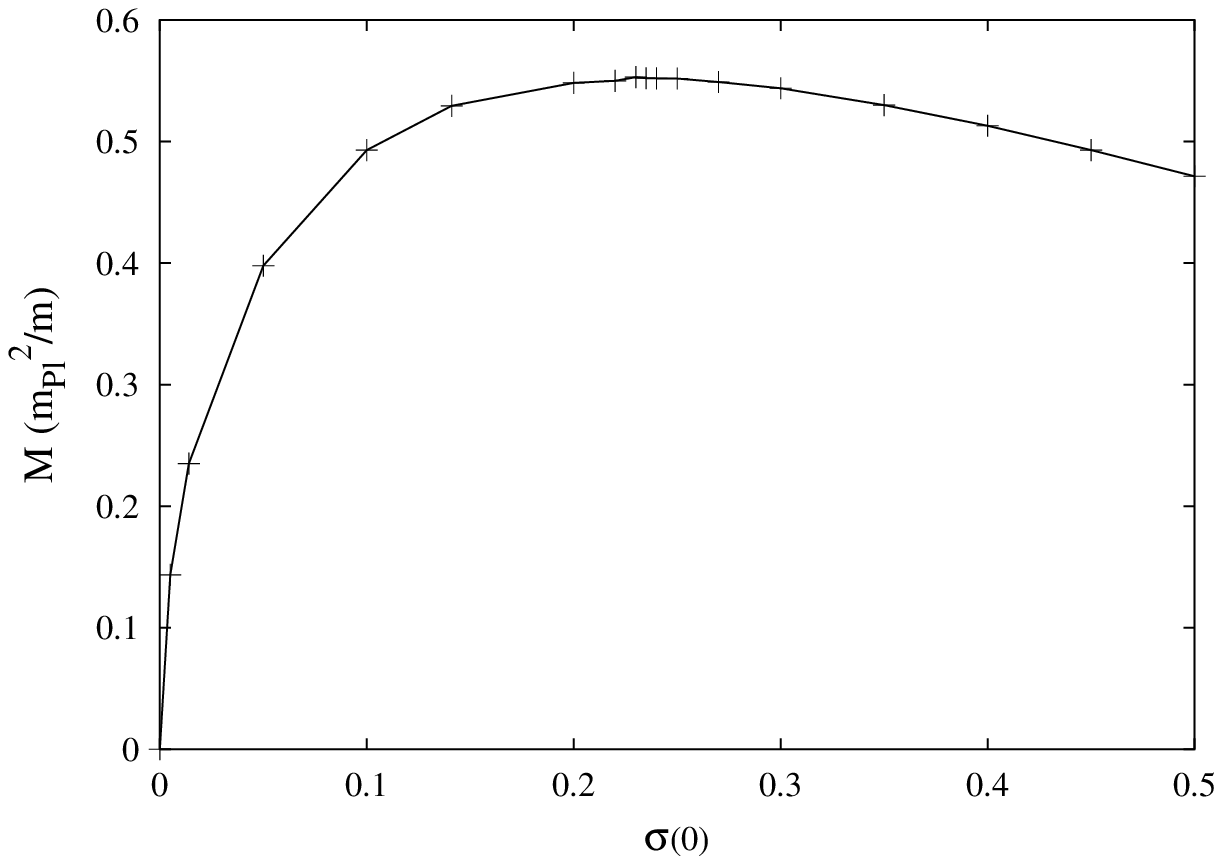}}
\caption{The energy density function $\rho_\Phi (\omega t,x)$ (see eq. \ref{rho}) for the oscillaton in Figs. \ref{fig:metos},\ref{fig:metfs} (top), and the mass observed at infinity (see eq. \ref{mass}) for different central values $\sigma(0)$ (bottom).}
\label{fig:rhos}
\end{figure}

The energy density for the oscillaton (\ref{tensor1}) is written

\begin{eqnarray}
\rho_\Phi (t,x) &=& \frac{1}{4\pi} m^2_{Pl} m^2 \left[ e^{-(\nu-\mu)} \sigma^2 \sin^2{(\omega t)} \right. \nonumber \\
&& \left. + e^{-(\nu+\mu)} {\sigma^\prime}^2 \cos^2{(\omega t)} + \sigma^2 \cos^2{(\omega t)} \right] \label{rho}
\end{eqnarray}

\noindent with $\mu,\nu$ as in (\ref{numu}). Its values for times $\omega t = 0, \pi/2$ are shown in Fig. \ref{fig:rhos}. It is easily noticed that $\rho_\Phi(0,x)=\rho_\Phi(\pi,x)$, and then the energy density may be expanded in a similar (albeit more involved) form as the metric coefficients (\ref{fexpg}). The same could also be done for the other quantities~(\ref{tensor2}-\ref{tensor4}).

Since the metric is asymptotically flat and static, in the sense that the oscillations of the metric are spatially confined (see Fig. \ref{fig:metos}), the mass seen by an observer at infinity may be calculated by the formula\cite{seidel90}

\begin{equation}
M_\Phi = (m^2_{Pl}/m) \lim_{x \rightarrow \infty} \frac{x}{2} \left(1-e^{-\nu-\mu} \right). \label{mass}
\end{equation}

\noindent The values found using this formula coincides with the ADM mass of oscillatons, but the former converges more rapidly and is more convenient from the numerical point of view.

The masses of different oscillatons are shown in Fig.~\ref{fig:rhos}. We can see that there is a maximum mass $M_{max} \simeq 0.5522 m^2_{Pl}/m$ with $\sigma_c(0) \simeq 0.235$. This critical value is important, since it was reported that oscillatons with lower central values than $\sigma_c(0)$ are stable\cite{seidel91}. The mass calculated in (\ref{mass}) is a constant and this implies that the masses observed at infinity are the same for all times. In fact, we see in Fig.~\ref{fig:metfs} that the oscillaton matches the Schwarzschild solution with the same mass\footnote{As it has to be according to Birkhoff's theorem\cite{wald}.}. 

As we mentioned before, the fundamental frequency is given by the asymptotic value $\exp{(\nu-\mu)} (\infty) = \Omega^{-2}$. Recalling that $\mu_0 (\infty)=-\nu_0 (\infty) \neq 0$ and taking into account the rapid convergence of function $\nu_0$ (see Fig.~\ref{fig:metfs}), we obtain the useful formula

\begin{equation}
\Omega = e^{-\nu_0 (\infty)}. \label{freq}
\end{equation}

\begin{figure}
\centerline{ \epsfysize=6cm \epsfbox{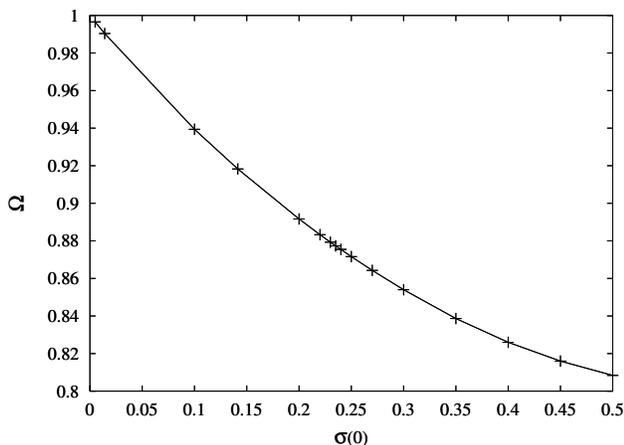}}
\caption{The fundamental frequencies $\Omega$ for the same cases shown in Fig.~\ref{fig:rhos}.}
\label{fig:omegas}
\end{figure}

The fundamental frequencies of oscillatons are given in Fig.~\ref{fig:omegas}. Observe that more massive oscillatons oscillate with smaller frequencies.

To finish this section, some lines about the approximation (\ref{sphi}). The neglected terms in eqs. (\ref{e2f2}-\ref{e3f2}) are proportional to $I_3(x \sigma \sigma^\prime)$ and $I_3(2 \mu_1)$, then the approximation is good enough if $(|x \sigma \sigma^\prime|,\, 2|\mu_1|) < 1$, or equivalently, $\sigma(0)< 0.5$. In this way, the differential equations contain only the leading terms of each expansion. But this is not exactly true for the KG eq. (\ref{kgf}), because a term proportional to $I_0(2 \mu_1) \cos(3 \omega t) \sim {\cal O}(1)$ was neglected. This term is related to the damping term in eq. (\ref{kg}) mentioned above. Thus, some physical information of the system is lost when taking just eq. (\ref{sphi}), since we should have included a time-dependent damping term.

An easy manner to address this problem is to take more terms in the expansion of the scalar field (\ref{sphi}), and then our ignorance on the damping term is transferred to the $3\omega t$-term. In doing so, the same situation will appear with the latter and we will need a $5\omega t$-term. In principle, the procedure have to be continued {\it ad infinitum}. However, the Fourier series converges rapidly\cite{seidel91,becerril} since the damping information transferred to the higher order terms becomes smaller in each step. 

The effect of this damping term on the stability of oscillatons can not be treated with the methods presented in this paper. But, we should mention that, using a full relativistic dynamical evolution of Einstein equations, it has been proved that oscillatons are stable configurations\cite{seidel91,seidel94}.

\section{The stationary limit procedure}
From the last paragraph, we can say that eqs. (\ref{e2f1}-\ref{e1f}) are a first and good approximation for solving oscillatons. Furthermore, the equations are simple enough that we can take a closer look at them and find in which form oscillatons and boson stars may be related.

If we take the weak field limit $\sigma(0) \ll 1$ (in consequence also $|\mu_1| \ll 1$), eqs. (\ref{e2f1}-\ref{kgf}) may be written as

\begin{eqnarray}
\nu^\prime_0 &=& x \left[ e^{2 \mu_0} \sigma^2 + {\sigma^\prime}^2 \right], \label{e2b} \\
\mu^\prime_0 &=& \frac{1}{x} \left[ 1 + e^{\nu_0+\mu_0} \left( x^2 \sigma^2 - 1 \right) \right], \label{e3b} \\
\sigma^{\prime \prime} &=& - \sigma^\prime \left( \frac{2}{x} - \mu^\prime_0 \right)+ \left( e^{\nu_0+\mu_0} - e^{2\mu_0} \right) \sigma. \label{kgb} \\
\mu^\prime_1 &=& e^{\nu_0+\mu_0} \sigma \left( x \sigma-\sigma^\prime \right), \label{e3b2}
\end{eqnarray}

\noindent where we have considered that $I_0(z) \sim {\mathcal O}(1)$ and $I_1(z)\sim {\mathcal O}(z/2)$ for $z \ll 1$. The time-independent parts of the oscillaton, eqs. (\ref{e2b}-\ref{kgb}) separates from the time-dependent ones, eqs. (\ref{e1f},\ref{e3b2}).

Eqs.~(\ref{e2b}-\ref{e3b2}) can be read directly from eqs. (\ref{e2}-\ref{kg}). Suppose that the scalar potential and its first derivative can be expanded in a Fourier series by taking eq. (\ref{sphi})

\begin{eqnarray}
V(\Phi) &=& V_0 (\sigma) + V_2 (\sigma) \cos{(2\omega t)} + ...\label{vf1} \\
V^\prime (\Phi) &=& V_1 (\sigma) \cos{(\omega t)} + ... \label{vf2}
\end{eqnarray}

\noindent This is possible if the scalar potential is even. Then, using again the dimensionless variables (\ref{dimenless}), eqs.  (\ref{e2b}-\ref{e3b2}) are just

\begin{eqnarray}
\nu^\prime_0 &=& x \left[ e^{2 \mu_0} \sigma^2 + {\sigma^\prime}^2 \right], \label{slimit1} \\
\mu^\prime_0 &=& \frac{1}{x} \left\{ 1 + e^{\nu_0+\mu_0} \left[ x^2 \kappa_0 V_0(\sigma)/m^2 - 1 \right] \right\}, \\
\sigma^{\prime \prime} &=& - \sigma^\prime \left( \frac{2}{x} - \mu^\prime_0 \right)+ e^{\nu_0+\mu_0} \sqrt{\kappa_0} V_1(\sigma)/(2m^2) - e^{2\mu_0} \sigma. \\
\mu^\prime_1 &=& e^{\nu_0+\mu_0} \left[ x \kappa_0 V_2(\sigma) /m^2- \sigma \sigma^\prime \right], \label{slimit2} 
\end{eqnarray}

\noindent while eq. (\ref{e1f}) remains the same.

The lowest order of approximation to the final result of the oscillaton, that in which the time-dependent and time-independent parts are separated, can be written using a harmonically time-dependent real scalar field and expanding the scalar potential in a Fourier series. I shall call this as the stationary limit procedure, in order to distinguish it from the treatment of section II. However, it must be stressed that this procedure will only work for potentials with a minimum. Otherwise, the assumption of coherent scalar oscillation does not make any sense. 

\begin{figure}
\centerline{ \epsfysize=6cm \epsfbox{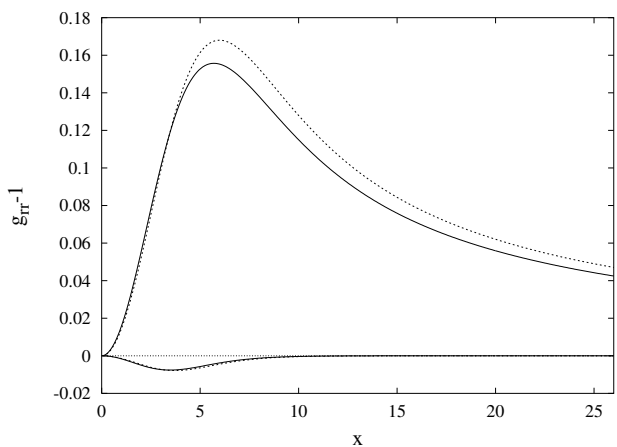}}
\centerline{ \epsfysize=6cm \epsfbox{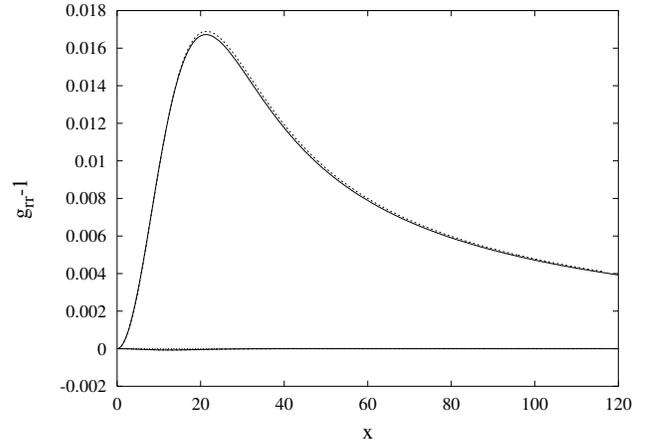}}
\caption{Comparison of the numerical results for the metric coefficient $g_{rr}-1$ for a quadratic oscillaton obtained with the method of sec. II (solid lines) and with the stationary limit (boson star-like) procedure (dashed lines), for the central values $\sigma(0)=0.2/\sqrt{2}$ (top), $0.02/\sqrt{2}$ (bottom). The curves represent ${g_{rr}}_0-1,{g_{rr}}_2$ as in Fig.~\ref{fig:metos}.}
\label{fig:bos}
\end{figure}

In Fig. \ref{fig:bos}, we show the numerical results for the metric coefficient $g_{rr}-1$ for an oscillaton using eqs. (\ref{e2f1}-\ref{e1f}) and for the boson star-like solutions to eqs. (\ref{e1f}, \ref{e2b}-\ref{kgb}), with the same central values $\sigma(0)=0.2/\sqrt{2},0.02/\sqrt{2}$. The adjusted values of $\mu_0 (0),\mu_1(0)$ for both cases nearly coincide, being almost indistinguishable for the weaker field.

A useful point about the stationary limit procedure is that, for the case of a quadratic scalar potential, eqs. (\ref{e2b}-\ref{kgb}) are of the same form than those worked for complex scalar field boson stars\cite{pang,seidel90,diego}.  Indeed, the central value $\sigma(0)$ is also the central value for the equivalent boson star.

A complete solution for oscillatons should have more Fourier terms in~(\ref{sphi}), that is, we should add more field contributions\cite{seidel91,becerril}. In general, the curves for an oscillaton are (a little bit) smaller than the those of a boson star represented in this case by the stationary limit procedure. This is a general fact since, unlike boson stars, oscillatons have an extra damping term in the KG equation~(\ref{kg}). Then, it would be found that oscillatons are in between boson stars (stationary limit procedure) and the results found in the previous section. 

However, the boson star-like equations are easier than a full treatment in Fourier series. Thus, the stationary limit procedure can be used to determine an upper limit for oscillatons and to provide, at least, order of magnitude results of interesting quantities. The results become exact when dealing with weaker fields. From now on, these are the general criterions we will take when working with the stationary limit procedure (see next section).

For instance, the maximum mass for a (0-node) boson star $M_{BS,c} = 0.633 m^2_{Pl}/m$ is reached when $\sigma(0)=0.268$\cite{seidel90}, while the maximum mass for a (0-node) oscillaton is $M_{\Phi,c} = 0.5522 m^2_{Pl}/m$ with $\sigma(0) \simeq 0.235$, see Fig. \ref{fig:rhos}. As we said before, the value of the maximum mass given by the boson star-like equations is an upper limit and of the same order of magnitude than the exact value\cite{seidel91,becerril}.

An interesting point now arises. From Fig. \ref{fig:metos}, we observe that the metric for an oscillaton is static from $x=10$ up. From Fig. \ref{fig:bos}, we also observe that it is difficult to distinguish between an oscillaton and a boson star far from the center, and in the weak field limit, the results are almost the same. This result suggests that the oscillaton is some kind of boson star plus an intrinsic oscillation (albeit this is not exact for non-polynomial potentials, as we will see in section V). The two objects are different only for an observer near the center. This last fact would imply that oscillatons could compete with boson stars as astrophysical objects of interest, since some useful properties of boson stars arises in situations far away from the boson star\cite{diego}, but the picture becomes more interesting if we move closer to the origin.

Concerning the possible Newtonian limit for oscillatons, we can see that such limit can not be obtained as a post-Newtonian expansion, as it was first pointed out in\cite{seidel91}. The stationary limit procedure seems to suggest that the static parts of the metric could have a kind of Newtonian limit as boson stars\cite{pang,arbey,moroz}, but the intrinsic oscillation, the very imprint of oscillatons, will not disappear\cite{becerril}.

\section{Self-interacting oscillatons}
Another class of interesting oscillatons are those with self-interacting scalar fields. Since these cases are more involved than the quadratic one, we will consider them within the stationary limit procedure only. 

Thus, the aim of this section is to obtain the qualitative behavior and approximate (at least in the order of magnitude) solutions to the leading terms in the Fourier expansions for the oscillatons. As it was said before, the true solutions must be smaller\footnote{The extra damping term in the KG equation~(\ref{kg}) does not depend on the strength of the scalar potential, then the results must not be much smaller.} but of the same order of magnitude.

As we will see below, self-interacting scalar fields provides with a wider range of properties that could be also of astrophysical interest. 

\subsection{The quartic case: $V=(m^2/2) \Phi^2 + (g/4) \Phi^4$}

This kind of potential has been considered as a model of bosonic dark matter\cite{tkachev,peebles,bento}. In this case, we will also set the dimensionless parameter

\begin{equation}
\Lambda = \frac{3 g}{\kappa_0 m^2} , \label{lambda}
\end{equation}

\noindent with $g/4$ the coefficient of the quartic term. Note that this parameter is different than that used for boson stars\cite{colpi}, but it is still a rough estimation of the ratio between the quartic and quadratic interactions in the potential. 

A complete treatment for this case is beyond the purpose of this paper. However, the stationary limit procedure can still give us enough information about the leading terms of the oscillaton functions. Taking this limit~(\ref{slimit1}-\ref{slimit2}), the equations to be solved are

\begin{eqnarray}
\nu_1 &=& x \sigma \sigma^\prime - \mu_1, \label{qekg1} \\
\nu^\prime_0 &=& x \left[ e^{2 \mu_0} \sigma^2 + {\sigma^\prime}^2 \right], \\
\mu^\prime_0 &=& \frac{1}{x} \left\{ 1 + e^{\nu_0+\mu_0} \left[ x^2 \left( \sigma^2 + (\Lambda/2) \sigma^4 \right)-1 \right]\right\}, \\
\sigma^{\prime \prime} &=& - \sigma^\prime \left( \frac{2}{x} - \mu^\prime_0 \right) + e^{\nu_0+\mu_0} \left( \sigma + \Lambda \sigma^3 \right) - e^{2\mu_0} \sigma \\
\mu^\prime_1 &=& e^{\nu_0+\mu_0} \sigma \left[ x \sigma+(2/3)\Lambda x \sigma^3 -\sigma^\prime \right].  \label{qekg2}
\end{eqnarray}

The parameter (\ref{lambda}) makes the last equations correspond to the case of boson stars\cite{colpi,seidel98}, and then the similarity between oscillatons and boson stars is preserved at this coupling. Following a similar argument as in\cite{colpi}, the self-interaction may be ignored if $\Lambda \ll 1$. But if $\Lambda \gg 1$, the oscillatons obtained may differ significantly from the non-interacting ones.

\begin{figure}
\centerline{ \epsfysize=6cm \epsfbox{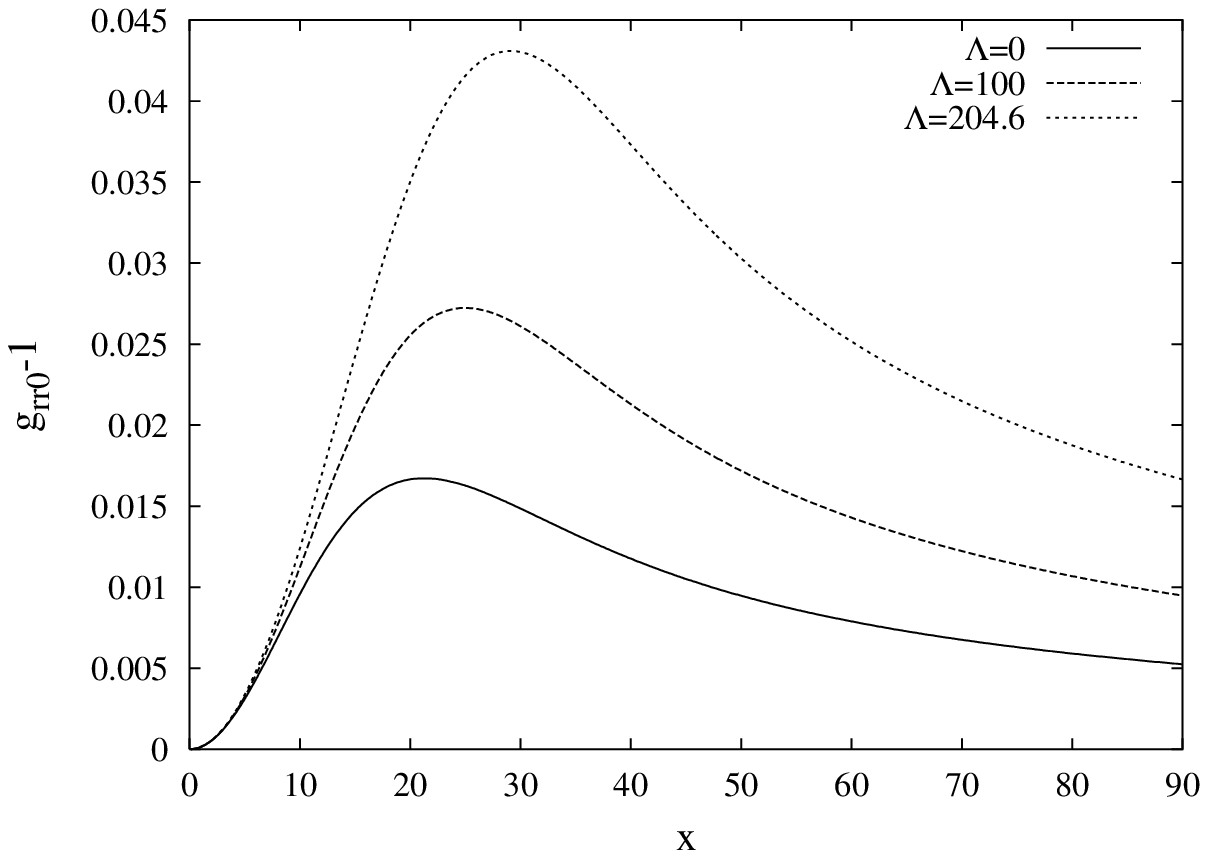}}
\centerline{ \epsfysize=6cm \epsfbox{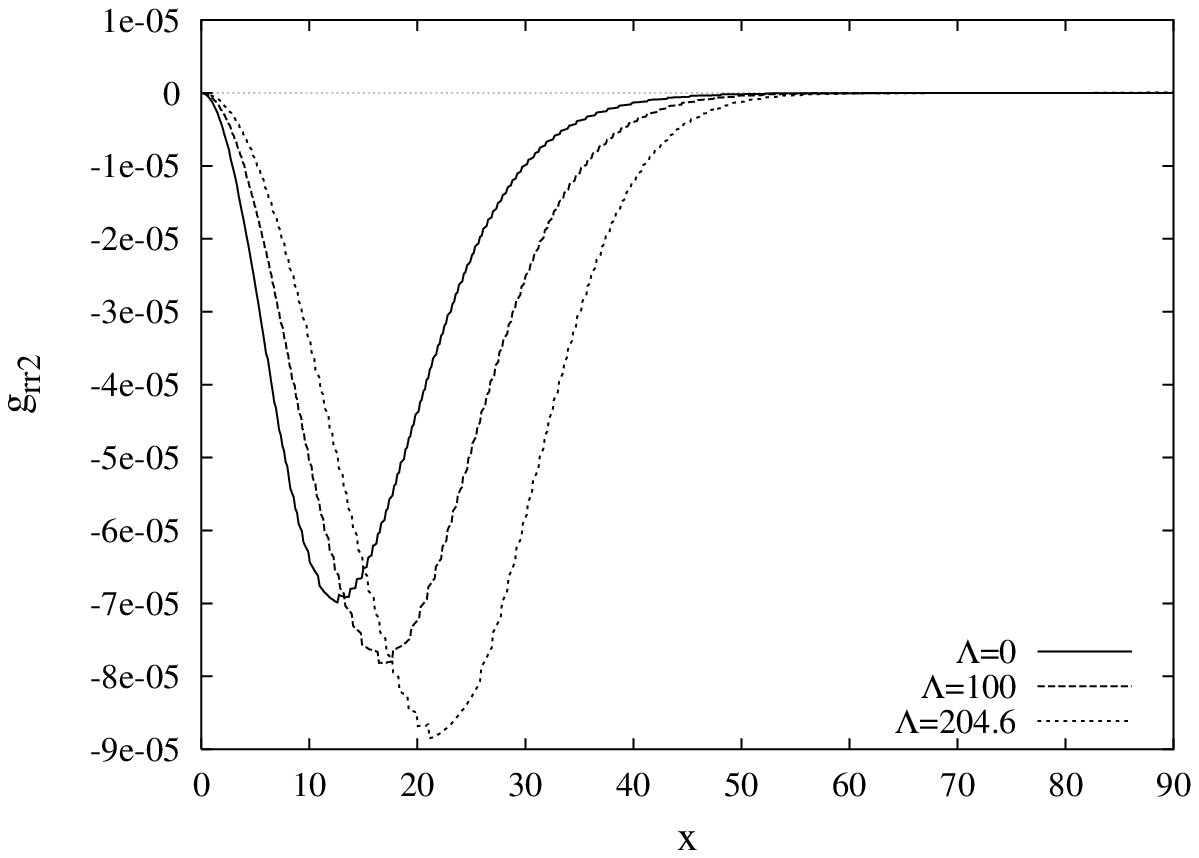}}
\caption{Comparison of the results for the metric coefficients ${g_{rr}}_0-1$ (top), ${g_{rr}}_2$ (bottom), for the case of an oscillaton with different quartic self-interactions within the stationary limit procedure. The central value is $\sigma(0)=0.02/\sqrt{2}$.}
\label{fig:osciself}
\end{figure}

Typical numerical results of an oscillaton with a quartic self-interaction within the stationary limit procedure are shown in Fig. \ref{fig:osciself}. As in the quadratic case, we can calculate the mass and frequency of those solutions: $M =\{0.736,\, 0.4225\}\, m^2_{Pl}/m$ and $\Omega=\{0.9787,\, 0.9857\}$ for  $\Lambda=\{204.6,\, 100\}$, respectively. These numbers can be compared with the $\Phi^2$ case, $M=0.235 \, m^2_{Pl}/m, \, \Omega= 0.9904$. The self-interacting oscillatons can be more massive and oscillate stronger but with a smaller frequency.

We can take advantage of the similarity with boson stars to approximately determine an upper limit for the maximum mass of an oscillaton with large-$\Lambda$. The result would be given by\cite{colpi}

\begin{equation}
M_{max} \simeq 0.2 \sqrt{\Lambda} m^2_{Pl}/m \label{maxm}
\end{equation}

\noindent where $\Lambda$ is given by~(\ref{lambda}).

Another interesting point has to do with the critical value $\sigma_c (0)$ at  which an oscillaton reaches the maximum mass. It is also important because oscillatons would be stable only for values $\sigma (0) \leq \sigma_c (0)$ as it is the case for boson stars\cite{seidel98}. 

Recalling the case of boson stars, the critical value becomes smaller for larger $\Lambda$. For instance, $\sigma_c (0) \simeq \{0.1,0.07\}$ for $\Lambda=\{100,204.6\}$ respectively, see Fig. 1 in\cite{colpi}. Moreover, for large-$\Lambda$, the critical value is approximately given by

\begin{equation}
\sigma_c (0) \simeq \Lambda^{-1/2} \label{sigmam}
\end{equation}

\noindent as suggested by Fig. 4 in\cite{colpi}.

This result can be taken as an upper limit of $\sigma_c (0)$ for oscillatons. That is, we can say that there must be stable oscillatons only for central values smaller then~(\ref{sigmam}). In both cases, boson stars and oscillatons, we need weaker fields for supporting stable self-interacting configurations.

Other useful information may be obtained from boson stars\cite{colpi,seidel98} and then approximately applied to oscillatons through the stationary limit procedure. In a similar way as in boson stars, quartic-oscillatons would provide a new range of configurations with more interesting properties than the quadratic ones.

\subsection{Non-polynomial potentials: $V(\Phi) = V_0 \left[ \cosh{(\lambda \sqrt{\kappa_0} \Phi)} -1 \right]$}

In a series of recent papers, our group have been investigated the hypothesis of scalar field dark matter in the Universe\cite{matos,matos2,luis1,luis2,luis3,matos3,matos4,matos5,lidsey}. It has been shown that a minimally coupled scalar field $\Phi$ endowed with a scalar potential

\begin{equation}
V(\Phi) = V_0 \left[ \cosh{(\lambda \sqrt{\kappa_0} \Phi)} -1 \right] \label{cosh}
\end{equation}

\noindent mimics quite well all known and desirable properties of cold dark matter at cosmological scales\cite{luis1,luis2}. Moreover, the free parameters of the potential $V_0$, $\lambda$ can be fixed from cosmological observations\cite{luis2}. The key point for this model is that the scalar potential possesses a minimum and that the scalar potential behaves as $\sim \Phi^2$ at late times in the evolution of the Universe. This scalar potential belongs to the family of scalar potentials of the form

\begin{equation}
V(\Phi) = \tilde{V}_0 \left[ \sinh{(\alpha \sqrt{\kappa_0} \Phi)}\right]^\beta \nonumber 
\end{equation}

\noindent which has been studied in quintessence\cite{sahni,luis4} and dark matter\cite{sahni,luis1,luis2,luis3,matos4,matos5}.

What kind of oscillatons would appear with~(\ref{cosh})?. First of all, the self-interaction is a very important issue\cite{luis3}. At first sight, we may take~(\ref{cosh}) with just a quartic coupling as in the previous section, with $g= \kappa_0 m^2 \lambda^2 /3!$ and the scalar field mass defined by $m^2 = \kappa_0 V_0 \lambda^2$. From eq. (\ref{lambda}), we have then that

\begin{equation}
\Lambda = \frac{1}{2} \lambda^2.
\end{equation}

For instance, taking a scalar dark matter model with a cosh potential\cite{luis1,luis2} cosmological observations suggest $\lambda = 20.28$ ($\Lambda \simeq 205.6$). In this particular case, we would have strong self-interacting oscillatons like those shown in Fig. \ref{fig:osciself}.

The full treatment of potential~(\ref{cosh}) is even more complicated than the quartic one. In order to have approximate information about the leading terms of the solution, let us take the stationary limit procedure (\ref{slimit1}-\ref{slimit2}). Considering that

\begin{eqnarray}
\left[ \cosh{(2 \lambda \sigma \cos{(\omega t)})}-1 \right] &=& \left( I_0(2 \lambda \sigma)-1 \right) \nonumber \\
&& + 2 I_2(2\lambda \sigma) \cos{(2\omega t)} + ..., \nonumber \\
\sinh{(2 \lambda \sigma \cos{(\omega t)})}   &=& 2 I_1(2 \lambda \sigma) \cos{(\omega t)} + ..., \nonumber
\end{eqnarray}

\noindent the Einstein-Klein-Gordon equations read

\begin{eqnarray}
\nu_1 &=& x \sigma \sigma^\prime - \mu_1, \label{cekg1} \\
\nu^\prime_0 &=& x \left[ e^{2 \mu_0} \sigma^2  + {\sigma^\prime}^2 \right] \\
\mu^\prime_0 &=& \frac{1}{x} \left\{ 1+ e^{\nu_0+\mu_0} \left[ \frac{x^2}{\lambda^2} \left( I_0(2 \lambda \sigma) -1 \right) - 1 \right] \right\} \\
\sigma^{\prime \prime} &=& - \sigma^\prime \left( \frac{2}{x} - \mu^\prime_0 \right) + \frac{1}{\lambda} e^{\nu_0+\mu_0} I_1(2 \lambda \sigma)- e^{2\mu_0} \sigma \\
\mu^\prime_1 &=& e^{\nu_0+\mu_0} \left[ \frac{2 x}{\lambda^2} I_2(2\lambda \sigma)-\sigma \sigma^\prime \right]. \label{cekg2}
\end{eqnarray}

It is worth to notice that the mass of the boson also sets the scale of the solution as in the cases before, albeit a more natural choice for this case could be the quantity $\sqrt{\kappa_0 V_0}$\cite{matos5}. The appearance of Bessel functions when expanding the scalar potential also shows that all couplings continue participating. Unfortunately, the case of a cosh-boson star has been treated before\cite{diego2} and then we see that the similarity among cosh-oscillatons and cosh-boson stars does not exist anymore.

Since we are mostly interested in stable cosh-oscillatons, we should first estimate $\sigma_c (0)$. From the $\Phi^4$-oscillatons we know that the stronger the interaction, the smaller the critical value $\sigma_c (0)$. Then, the value (\ref{sigmam}) must be an upper limit for the cosh case. Therefore, for cosh-oscillatons we obtain

\begin{equation}
\sigma_c (0) \simeq \sqrt{2} \lambda^{-1}.
\end{equation}

But, observe that for these values (and smaller ones) the cosh potential (\ref{cosh}) can be treated perturbatively. That is, eqs. (\ref{cekg1}-\ref{cekg2}) can be worked in the limit of eqs. (\ref{qekg1}-\ref{qekg2}). Therefore, as long as we consider only stable oscillatons, the cosh potential gives (almost exactly) the same results as the quartic potential does.

Going further, the maximum mass for the cosh oscillaton must be of order $M_{max} \simeq 0.1 \lambda m^2_{Pl}/m $ (see eq. (\ref{maxm})). This formula can be written in a more convenient form by taking the definition of the scalar field mass. Then, the maximum mass for a cosh oscillaton is approximately given by

\begin{equation}
M_{max} \simeq 0.1 \frac{m^2_{Pl}}{\sqrt{\kappa_0 V_0}},
\end{equation}

\noindent which in turn means a $\lambda$-independent value.

The validity of the last formula and the existence of stable cosh-oscillatons have been recently confirmed using a full numerical evolution of the Einstein-Klein-Gordon equations\cite{matos5}.

\section{Conclusions}

The simplest approximation for solving the coupled Einstein-Klein-Gordon equations for a spherically symmetric $\Phi^2$-oscillaton were presented, by taking an harmonic time-dependent scalar field. Non singular and asymptotically flat solutions are obtained taking a Fourier expansions of the differential equations. These solutions are also asymptotically static, so the metric oscillations are confined to the region of the oscillaton. Using this last fact, we could determine the mass and fundamental frequency of some configurations. The questions concerning the stability and formation of these oscillatons were answered based in previous publications.

We presented a method called the stationary limit procedure, which gives the lowest order of approximation to the leading terms in the Fourier expansions. In this limit, the differential equations for the oscillaton are boson star-like and the similarity between these objects becomes manifest. Taking this similarity, we could determine an upper value and an order of magnitude estimation for some critical parameters of oscillatons by comparing them with boson stars.

The cases of quartic and cosh self interactions need a more complicated numerical treatment. However, they were studied in the stationary limit procedure. For the quartic case, boson star-like differential equations arise again. Thus, the solutions given in the literature for boson stars apply for oscillatons, at least at the level of differential equations. In the non-polynomial case of a cosh scalar potential, the equations are not boson star-like, but the results are quite similar to the case of quartic self-interaction, as long as we refer only to stable configurations. About the stability of these oscillatons, we determined the central value $\sigma_c(0)$ at which the maximum mass is reached. In the quadratic case, there is evidence that the oscillaton is unstable for central values larger than $\sigma_c(0)$, and we assume the same for the self-interacting ones. Whether this really happens is still matter of current research and the results will be published elsewhere\cite{matos5,becerril}.

There are some issues to be investigated. First of all, it is desirable to have a more useful and complete numerical treatment to find the oscillaton configurations, one in which the non-linearities of the EKG equations could be minimized as much as possible\cite{becerril}. A similar step should be made to include self-interacting oscillatons. This would allow us to exactly determine the reliability of the stationary limit procedure.

As it was shown through the stationary limit procedure, it must be possible to find a kind of Newtonian-like limit of oscillatons. Apart from the simplicity and the appealing scaling properties that may appear (for boson stars and their Newtonian limit see\cite{pang,sin,arbey,moroz}), such limit would be also of cosmological and galactic interest, like in the case of Newtonian boson stars\cite{sin,arbey,jones}. Work in that direction is in progress\cite{becerril}.

Finally, scalar dark matter would provide of very interesting gravitationally bound objects like oscillatons. These relativistic systems, with no Newtonian counterpart, would appear from the gravitational collapse of a real scalar field, and their properties may be directly connected to cosmological parameters\cite{luis1,luis2,luis3,matos3,matos4,matos5,lidsey}. This additional reason makes oscillatons worth investigating.

\acknowledgments{I would like to thank Tonatiuh Matos, F. Siddhartha Guzm\'an and Ricardo Becerril for many helpful discussions. This work was supported by CONACyT, M\'{e}xico under grants 119259 and 34407-E.}


\end{multicols}

\end{document}